*Article*

# The van der Waals Hexaquark Chemical Potential in Dense Stellar Matter

Keith Andrew[1],*, Eric V. Steinfelds[1,2] and Kristopher A. Andrew[3]

1 Department of Physics and Astronomy, WKU Western Kentucky University Bowling Green, KY,42101, USA; eric.steinfelds@wku.edu (E.V.S.); kristopher.andrew@outlook.com (K.A.A.)
2 Department of Physics, Marquette University, Milwaukee, WI 53233, USA;
3 Department of Science, Schlarman Academy, Danville, IL 61832, USA
* Correspondence: keith.andrew@wku.edu

**Abstract:** We explore the chemical potential of a QCD-motivated van der Waals (VDW) phase change model for the six-quark color-singlet, strangeness S = −2 particle known as the hexaquark with quark content (uuddss). The hexaquark may have internal structure, indicated by short range correlations that allow for non-color-singlet diquark and triquark configurations whose interactions will change the magnitude of the chemical potential. In the multicomponent VDW Equation of State (EoS), the quark-quark particle interaction terms are sensitive to the QCD color factor, causing the pairing of these terms to give different interaction strengths for their respective contributions to the chemical potential. This results in a critical temperature near 163 MeV for the color-singlet states and tens of MeV below this for various mixed diquark and triquark states. The VDW chemical potential is also sensitive to the number density, leading to chemical potential isotherms that exhibit spinodal extrema, which also depend upon the internal hexaquark configurations. These extrema determine regions of metastability for the mixed states near the critical point. We use this chemical potential with the chemical potential modified TOV equations to investigate the properties of hexaquark formation in cold compact stellar cores in beta equilibrium. We find thresholds for hexaquark layers and changes in maximum mass values that are consistent with observations from high mass compact stellar objects such as PSR 09043 + 10 and GW 190814. In general, we find that the VDW-TOV model has an upper stability mass and radius bound for a chemical potential of 1340 MeV with a compactness of C~0.2.

## 1. Introduction

The thermodynamics of strongly interacting matter and the phase structure of QCD have been studied extensively from both a lattice perspective and from QCD-motivated phenomenological models. Many of these models have been bolstered by the rapidly growing accelerator measurements of the properties of quark gluon plasma and by astrophysical observations of compact dense stellar cores. However, the complexities and nonlinear dynamics of QCD have made it difficult to directly understand all of the properties of novel quark matter-bound states, such as tetraquark [1], pentaquark [2,3], and hexaquark [4] particles (note: authors also refer to the 6-quark particle as the sexaquark so the H symbol will not be confused with the Higgs). In the appropriate limits, such states can be modeled as loosely bound molecular quark states—diquarks with bound mesons, with color states $q_i \bar{q}_j \delta_{ij}$ or baryons, with color states $\varepsilon_{ijk} q_i q_j q_k$, as SU(3)$_c$ color-singlet states or more strongly bound compact single-particle states, or as less common hybrid states of quarks and gluons or pure gluonic states [5]. For states with more than three quarks, there exist more pairings that give color-singlet states, for example [6], the $\bar{3}_c \otimes 3_c = 1_c$ and the $6_c \otimes \bar{6}_c = 1_c$ result in a varied spectrum of states. The existence of such states indicates that a system such as a cold compact stellar core can consist of a mixture of quark clusters such as diquarks and triquarks [7] or as short length correlated groups in a particle such as a hexaquark, along with various color-singlet states. Various hexaquark flavor, isospin, spin, and angular momentum states have been studied, beginning with Jaffe examining a $J^P = 0^+$ dihyperon [8] with a focus on the d*(2380) $I(J^P) = 0(3^+)$ [9–11] and interest in the (uuddss) flavor-singlet, charge-neutral, even-parity, spin-zero boson with baryon number and strangeness B = 2, S = −2. This last case is especially interesting because, as noted by Farrar [12,13], it may have a long-life ground state, making it an interesting candidate for dark matter and impacting the internal structure of compact stellar cores. As the pressure increases towards the central region of the core, the number of particles in each state will change and the central core region at the highest pressure can undergo a phase transition from a bound to an unbound quark gaseous state. Several authors have investigated this phase change for a cold neutron star model that is charge neutral and in beta equilibrium utilizing a system of quark clusters [14], quasiparticles [15], and quark drops [16] as partial intermediate states as the system approaches a free-quark gaseous state. One widely used model to investigate this behavior is a simple analytical model based upon the multicomponent van der Waals (VDW) equation of state (EoS), which incorporates particle species' chemical potentials to accommodate changing particle numbers and includes a first-order phase transition. Such a generalized VDW model focused on dense fluids [17] was further developed by Vera [18] and extended the development of the Prigogine [19]-Flory [20]-Patterson [21] theory. For dense matter, the general VDW partition function and statistical method of Eu [22] with the multicomponent partition function method of Keffer [23], as developed by Vovchenko [24–27], has the advantages of including the excluded particle volume, incorporating attractive and repulsive interactions, exhibiting a first-order phase transition, including multicomponent mixtures, showing binodal and spinodal behavior, having a well-defined chemical potential, and exhibiting a critical point. VDW-based models have become an important way to gain insight into the hadronic deconfining phase transition [28–30] and as a model for a hadronic gaseous state [31]. Here we will apply the VDW EoS to a system of hexaquarks where the hexaquarks can have different internal structures consisting of diquark and triquark states [32]. The different binding strengths given by the color factors are represented by the VDW mixing parameters; the multicomponent VDW equations are used to analyze a system with combinations of the various hexaquark states. For example, we can analyze the properties of a hexaquark fluid consisting of hexaquarks that have a three-diquark internal structure. Or we can have a two-component fluid consisting of one component made from hexaquarks with a two-triquark substructure and the other component consisting of hexaquarks with three diquarks. While the VDW equations allow us to find the chemical potential and critical point for each fluid, the multicomponent VDW equations allow us to find the chemical potential and critical point for the mixture.

A potential arena where the impact of hexaquark internal structure, critical points, chemical potentials, and phase change phenomena could take place, and be constrained by observation, is in the dense core of a neutron star or a possible quark star [33]. As the observational data have become more robust and refined, more detailed models have emerged to help understand the varied mechanisms at play in dense QCD matter. Several models are gaining support from the observations of quark–gluon plasma, QGP, demonstrating the existence of a high-temperature, low-chemical-potential state of unconfined quarks, as seen at the SPS [34], LHC [35], and RHIC [36] laboratories. These experiments give a transition temperature near 155 MeV and an energy density near 0.8 GeV/fm$^3$ [37]. Stellar cores represent systems of high baryon chemical potential, with high density and pressure which might be capable of exhibiting a deconfinement phase transition at high pressure as noted by Baym, et al. [38]. Isolated nonaccreting neutron stars are cold, less than ~0.1 MeV, and after a few hundred years can be nearly isothermal [39]. For larger masses and higher pressures, a transition to a quark star can occur which may have spin or tidal deformation [40], strange quarks [41,42], magnetic field effects [43], or color superconductivity [44]. For the known transition temperature, there is a baryon chemical potential, or, equivalently, a density or pressure, where the phase transition will occur which can be described by the EoS [45] and, when considering the case of an isotropic density and pressure as source terms, can be described by the TOV equations. As outlined in Baym [46], the constraints of charge neutrality and beta equilibrium can be used to estimate the chemical potential within the context of the MIT bag model, while at high temperatures, T > 1 MeV, the matter is out of beta equilibrium [47]. Charge neutrality for particle number density $n_j$, mass density $\rho_f$, and electric charge $q_j$ for particle type, j, or flavor, f, can be expressed as $\sum q_f n_f = q_e n_e$ and $\rho_B = \rho_u + \rho_d + \rho_s$. The three-light quark flavors, changing weak interaction equilibrium conditions from the quark interactions, given by $d \leftrightarrow u + e^- + \bar{\nu}_e$, $s \leftrightarrow u + e^- + \bar{\nu}_e$, and $s + u \leftrightarrow d + u$, constrain the chemical potentials. We consider the late-time case, in which the neutrinos and antineutrinos have exited the collapsed core on a time scale that is short compared to the long-term cooling time, effectively causing their chemical potentials to vanish to establish beta equilibrium, resulting in $\mu_d = \mu_u + \mu_{e^-}$ and $\mu_s = \mu_d$. Then, the pressure and energy density ε are given by $P = -\partial U / \partial V = n^2 \partial(\varepsilon/n)/\partial n = n\mu - \varepsilon$, where the chemical potential is $\mu = \partial\varepsilon/\partial n$ and n is the number density where the nuclear saturation density is $n_{sat}$ ~ 0.16 fm$^{-3}$. For compact stars, these models allow for comparison to the MIT bag model [48] and the modified MIT bag model [49–50] for the pressure, P, energy density, ε, and chemical potential, μ, with the bag constant, *B*, expressed as $3P = (\varepsilon - 4B) = \varepsilon - 4(3\mu^4/4\pi)$, relating the baryon chemical potential to the bag constant, which is subject to the Franzon [51] constraint by stability requirements in neutron star models: 30 MeV/fm$^3$ < B < 75.5 MeV/fm$^3$. The chemical potential for multiquark particles is given by μ = m+$E_F$, where $E_F$ is the Fermi energy. For a system of noninteracting fermions, the Fermi energy is given by $E_F = (3^{2/3} \pi^{4/3} 2^{-1/3}) \hbar^2 (n/(2s+1))^{2/3}/m$ for the reduced Planck constant, $\hbar$, spin, s, number density, n, and mass, m [52,53]. For hexaquark states consisting of quarks, diquarks, and triquarks, which can have spin states s = 0, 1/2, 1, 3/2, the Fermi energy is in the range of 55 –500 MeV and chemical potentials are in the range of 850 MeV–2100 MeV [54,55]. Knowledge of the baryon chemical potential and pressure in the core provides an important method for identifying a deconfining phase transition. Within the context of the Maxwell construction [56], this occurs when the hadronic and quark pressures and chemical potentials of quarks and leptons [57] are equal: $P_h = P_q$ and $\mu_h = \mu_q$, where $\mu_q = 3\left(\sum_{quarks}\mu_q n_q + \sum_{leptons}\mu_l n_l\right)/\sum_{quarks} n_q$ and $\mu_h = \left(\sum_{hadrons}\mu_h n_h + \sum_{leptons}\mu_l n_l\right)/\sum_{hadrons} n_h$ for the hadron, h, and lepton, l, labels for the chemical potentials and number densities. We only consider the case after the neutrinos have escaped; however, a more careful treatment by Dexheimer considers the protostar case with trapped neutrinos [58].

Here we will develop this model to find the range of chemical potentials of the hexaquark that can exist in a compact core within the Franzon stability range [59,60]. We will first introduce the VDW model and match the parameters to the quark interactions applicable to the determination of the chemical potential of the hexaquark. Using the multicomponent VDW equation, we examine the differences in the chemical potential that result from the molecular and independent constituent models of the hexaquark substructure. Values from the SHM [61] at RHIC [62] and ALICE [63] are matched with lattice values [64] to determine the functional form of the temperature dependent chemical potentials. We then examine the variation in chemical potential exhibited in a dense stellar core by solving the TOV system for the chemical potential. We use natural units where the Boltzmann constant, the speed of light, and Newton's gravitational constant are set to unity.

## 2. van der Waals Model Chemical Potential in a Hadronic Mixture

In this application of the VDW EoS, we consider a uniform state of bound quark clusters that can undergo a phase transition to free quarks, as performed by Zakout for the quark gluon plasma [65]. In the multicomponent VDW EoS, the system can consist of several different components which correspond to different types of clusters; here we are limiting the model to clusters that yield a hexaquark, i.e., each system consists of hexaquarks but the underlying hexaquark structure is governed by different short-range correlations giving different color factors which are modeled in the VDW mixing factors. These can consist of diquark and triquark clusters, each of which form hexaquarks, that can be mixed with a hexaquark with no internal structure. This system can then be viewed as a multicomponent fluid where each component is described by its own chemical potential. Following the statistical development of VDW EoS, we consider the multicomponent, $N_c$, van der Waals partition function given by

$$Z_{vdw}(N_i, V, T) = \prod_{i=1}^{N_c} \frac{1}{N_i!} \left( \frac{V - \sum_{j=1}^{N_c} N_j b_j}{\Lambda_i^3} \right)^{N_i} \exp\left( \frac{N_i}{VkT} \sum_{j=1}^{N_c} N_j a_{ij} \right) \tag{1}$$

Where $b_j$ is the van der Waals effective volume of the $j^{th}$ particle of number $N_j$, the thermal de Broglie wavelength is

$$\Lambda_i = \sqrt{\frac{1}{2\pi m_i T}} \tag{2}$$

and the van der Waals interaction parameter is $a_{ij}$. We adopt the notation of $a_{ii} = a_i$, noting that the $a_{ij}$ is symmetric ($a_{ij} = a_{ji}$) and that there is a mixing rule, $a_{ij} = \sqrt{a_i a_j}(1 - k_{ij})$, where $k_{ij}$ is a mixing parameter that is used to account for the color factor interaction differences between the strength of the diquark color interaction for non-singlet states and for color-singlet states. Using Sterling's approximation, the pressure is

$$p = T\left(\frac{\partial \ln Z}{\partial V}\right)_{N,T} = T\sum_{i=1}^{N_c} \left[ \frac{N_i}{V - \sum_{j=1}^{N_c} N_j b_j} - \frac{N_i}{V^2 T} \sum_{j=1}^{N_c} N_j a_{ij} \right] \tag{3}$$

The pressure expression is the equation of state, EoS, for our system. In terms of the single component number density, $n = N/V$, Equation (3) can be used to find the VDW speed of sound as

$$c_s^2 = \left(\frac{1}{m}\frac{\partial P}{\partial n}\right) = \frac{T}{m(bn-1)^2} - \frac{2an}{m} \tag{4}$$

which in the limit of vanishing VDW constants gives the ideal gas law value of $T/m$. The critical point of the phase diagram can be found by solving the system of equations,

$$P = P(V,T,N), \quad \frac{\partial P}{\partial V} = 0, \quad \frac{\partial^2 P}{\partial V^2} = 0$$

$$P_c = \frac{\left(\sum_{i=1}^{N_c} N_i \sum_{j=1}^{N_c} N_j a_{ij}\right)}{27\left(\sum_{j=1}^{N_c} b_j N_j\right)^2}, \quad T_c = \frac{8\left(\sum_{i=1}^{N_c} N_i \sum_{j=1}^{N_c} N_j a_{ij}\right)}{27k\left(\sum_{i=1}^{N_c} N_i\right)\left(\sum_{j=1}^{N_c} b_j N_j\right)}, \quad V_c = 3\sum_{j=1}^{N_c} b_j N_j \quad (5)$$

for the critical values $P_c$, $V_c$, and $T_c$, while the resulting chemical potential for the $i^{\text{th}}$ species is

$$\mu_i = -T\left(\frac{\partial \ln Z}{\partial N_i}\right)_{T,V,N_{j\neq i}} = T\ln N_i - T\left[\ln\left(\frac{V-\sum_{j=1}^{N_c} N_j b_j}{\Lambda_i^3}\right) - \frac{b_i}{V-\sum_{j=1}^{N_c} N_j b_j}\sum_{j=1}^{N_c} N_j + \frac{2}{VT}\sum_{j=1}^{N_c} N_j a_{ij}\right] \quad (6)$$

When the van der Waals volume correction is small compared to the total volume, the logarithm term can be expanded as a power series and regrouped to express the chemical potential as the sum of a term that is independent of the van der Waals constants and a term with the explicit dependence on the van der Waals constants

$$\mu_i = \mu_{i_o} + \mu_{i_{ab}} = \mu_{i_o} - T\left[\sum_{k=1}^{\infty}\frac{1}{k}\left(\sum_{j=1}^{N_c}\frac{N_j b_j}{V}\right)^k - \frac{b_i}{V-\sum_{j=1}^{N_c} N_j b_j}\sum_{j=1}^{N_c} N_j + \frac{2}{VT}\sum_{j=1}^{N_c} N_j a_{ij}\right] \quad (7)$$

$$\mu_{i_o} = T\ln\left(\frac{N_i \Lambda_i^3}{V}\right) = T\ln\left(n_i \Lambda_i^3\right) \quad ,$$

where, for a single component small particle volume, $b_j/V \ll 1$, with $i = j = k = 1$, the chemical potential in terms of the number density, $n_j = N_j/V$, simplifies to

$$\frac{\mu_1}{T} = \ln\left(n_1 \Lambda_1^3\right) - \left[\frac{2n_1 a_{11}}{T} - \frac{n_1^2 b_1^2}{1 - n_1 b_1}\right] \quad (8)$$

This result can now be used with the measured values of the quark chemical potentials to determine the van der Waals constants and the chemical potential in dense matter, such as the central core of a compact star. For $N_c$ particles in a system with equal number densities for each particle, $n = n_1 = n_2$, and symmetric interaction mixing for color factors, $k_{ii} = k_{ji}$, at equilibrium, the total chemical potential is

$$\mu = \sum_{i=1}^{N_c} \mu_i$$

$$\mu_i = \mu_{i_o} - T\left[n\sum_{j=1}^{N_c} b_j - \frac{2nb_i}{1 - n\sum_{j=1}^{N_c} b_j}\right] - 2\sum_{j=1}^{N_c} n_j \left(a_i a_j\right)^{1/2}\left(1 - k_{ij}\right) \quad (9)$$

In the VDW model, the chemical potential is singular at the phase transition where the effective volume of the constituents approaches the volume of the object when the number density is sufficiently large; such a density can arise in a compact stellar core during collapse. We use the chemical potentials for baryon number, isospin, and strangeness from SHM and lattice models to find the VDW constants consistent with Equation (8) for the three lightest quarks and the mass values from the PDG review of particle properties [66]. These values are then used to determine the chemical potentials and VDW constants for the systems of combined quarks forming particles with net-color or color-singlet states, the diquark, and triquark states. These form the building blocks for the hexaquark color-singlet states containing six quarks in the VDW model.

The composite particle states are constructed using the quark values from Table 1 with color factors to match the resulting state and find the chemical potential. These values represent the scalar S ground states which are nearly 220 MeV below their axial counterparts and are given in Table 2 [67,68].

**Table 1.** Quark VDW constants from baryon, isospin, and strangeness chemical potentials.

| Particle | m [MeV] | a [GeV$^{-2}$] | b [GeV$^{-3}$] |
| --- | --- | --- | --- |
| Up | 2.2 | 0.0011 | 0.00201 |
| Down | 4.7 | 0.0012 | 0.00217 |
| Strange | 93 | 0.0037 | 0.00683 |

**Table 2.** van der Waals constants for the diquark and triquark, as well as different representations of the internal structure of the hexaquark.

| Particle | m [MeV] | a [GeV$^{-2}$] | b [GeV$^{-3}$] |
| --- | --- | --- | --- |
| Diquark (ud) | 509 | 0.0098 | 0.0182 |
| Diquark (ds) | 698 | 0.0137 | 0.0251 |
| Triquark (uds) | 2077 | 0.0419 | 0.0748 |
| Hexaquark (uuddss) | 2110 | 0.0472 | 0.0839 |
| 3-diquarks (ud)(su)(ds) | 1883 | 0.0518 | 0.0914 |
| 2-triquarks (uds)(uds) | 2324 | 0.0566 | 0.0987 |

These values can then be used with Equations (6) and (7) to express the chemical potentials as functions of temperature; plots for special cases are given in Figure 1 below.

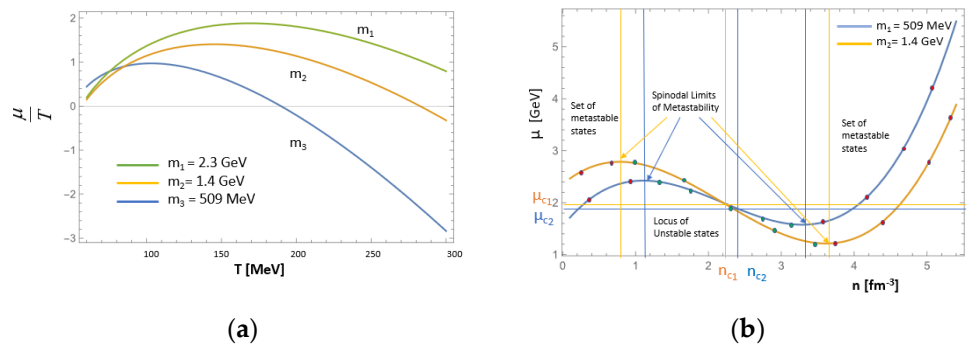

(**a**)        (**b**)

**Figure 1.** (**a**) The chemical potential in terms of the dimensionless thermodynamic variable μ/T as a function of temperature with increasing mass values; using our extrema and midpoint, the masses are: $m_1$ = 2.3 GeV, $m_2$ = 1.4 GeV, and $m_3$ = 509 MeV and (**b**) chemical potential isotherms near the critical point exhibt the spinodal and metastable regions characteristic of a VDW EoS for two different mass values. The spinodal points are the local extrema of each curve.

In our analysis, the chemical potential, which depends upon $b_2$, needs to be real-valued, and this provides an additional constraint on Equation (6) giving a relationship between a, n, T, and the chemical potential

$$\frac{-4T + \mu_1 T - T^2 \ln(n_1 \Lambda_1^3)}{2n_1} \geq a \geq \frac{\mu_1 T - T^2 \ln(n_1 \Lambda_1^3)}{2n_1} \; . \tag{10}$$

These results can be applied to an environment where the Fermi energy is on the order of the chemical potential to explore shifts in the chemical potential, such as in a dense stellar core, which can be modeled using the TOV equations. Near the critical point, the expressions for the pressure and chemical potential can be simplified by expanding about the critical point, where $\mu_c$ and $n_c$ denote the chemical potential and density evaluated at the critical point, and using the dimensionless density ratio, z, resulting in the cubic expressions

$$z = \frac{n - n_c}{n_c}$$

$$\mu = \mu_c + \frac{9 T_c n_c}{4} z \left[ \frac{T}{T_c} - 1 + \frac{1}{4} z^2 \right] \tag{11}$$

$$\left(\frac{p - p_c}{p_c}\right)(2 - z) = 8\left(\frac{T}{T_c} - 1\right)(1 + z) + 3z^3$$

Using Equation (11), the pressure near the critical point can be expressed as a function of the chemical potential

$$p = \frac{p_c n \left(32 \mu_c n_c^2 - 32 n_c^2 \mu - 9\left(n^3 - 6 n_c n^2 + 11 n_c^2 n - 6 n_c^3\right) T_c\right)}{9(n - 3n_c)(n - n_c) n_c^2 T_c} \tag{12}$$

which is monotonically increasing up to the neighborhood of the critical point in a fashion similar to the MIT bag model. The extremum behavior of the chemical potential allows us to identify regions of metastability for the system in the neighborhood of the critical temperature. The chemical potential local minimum and maximum correspond to the spinodal limits of metastability while the phase transition is taking place. For the isotherms with $T < T_c$, the Maxwell construction replaces the equal area regions above and below the Maxwell constant pressure line during the mixed-phase transition in the phase diagram. It is in this region where condensing nanoclusters will form and coexist with the vapor phase; this does not shift the critical points used here where the isotherm of interest is along $T = T_c$ [69]. These points are located at

$$\frac{\partial \mu}{\partial n} = 0$$

$$n = n_c \left(1 \pm \frac{2}{\sqrt{3}} \sqrt{(1 - T/T_c)}\right) \tag{13}$$

For equal particle numbers of each component, the two-component critical temperature can be expressed as

$$T_{C-2} = \frac{8\left(a_1(1 - k_{11}) + 2\sqrt{a_1 a_2}(1 - k_{12}) + a_2(1 - k_{22})\right)}{27(b_1 + b_2)} \tag{14}$$

which can be used to evaluate the critical chemical potential. The VDW constants can be used to determine the VDW critical temperature and chemical potential using Equations (4) and (10), where we consider the mixed states of the diquark and dibaryon hexaquark substructures and examples of mixed diquark-triquark-hexaquark combinations as shown in Table 3..

**Table 3.** The critical values for the temperature from Equation (4) and chemical potential at the critical temperature for different hexaquark internal structures.

| Hexaquark Structure | $T_c$ [MeV] | $P_c$ [MeV/fm$^3$] | $n_c$ [fm$^{-3}$] | $\mu_c$ [MeV] |
|---|---|---|---|---|
| Hexaquark | 166.7 | 429.1 | 2.29 | 2703.9 |
| 3- diquarks | 167.9 | 396.8 | 2.11 | 2720.5 |
| 2- triquarks | 169.9 | 371.8 | 1.95 | 2667.8 |
| Mixed: Diquark-hexaquark: $k_{12}=0$, mixing 1:1 | 151.9 | 602.0 | 3.54 | 1125.8 |
| $k_{12}=0$, mixing 2:1 | 150.0 | 725.0 | 4.32 | 1881.2 |
| $k_{12}=0.5=k_{21}$, mixing 1:1 | 117.3 | 465.0 | 3.54 | 924.0 |
| $k_{12}=0.5=k_{21}$, mixing 2:1 | 112.6 | 544.0 | 4.31 | 1465.8 |
| Mixed Triquark-hexaquark: $k_{12}=0$, mixing 1:1 | 166.2 | 452.4 | 2.43 | 811.7 |
| $k_{12}=0=k_{21}$, mixing 2:1 | 166.1 | 460.9 | 2.47 | 858.7 |
| $k_{12}=0.5=k_{21}$, mixing 1:1 | 124.7 | 339.4 | 2.43 | 669.8 |
| $k_{12}=0.5=k_{21}$, mixing 2:1 | 128.5 | 356.5 | 2.48 | 714.3 |

The critical values obey the VDW compressibility factor rule that the term $P_c/(n_c T_c)$ is a constant at the critical point. It is useful to compare these results to the free quark MIT bag model. If we denote the bag constant by B, then the pressure, p, the baryon density, $\varrho_B$, and the speed of sound, $c_s$, are given by:

$$p = \frac{1}{3}\left(\frac{3}{2}\right)^{7/3} \pi^{2/3} \rho_B^{4/3} - B, \quad \rho_B = \frac{1}{3}(\rho_u + \rho_d + \rho_s), \quad c_s = \sqrt{\frac{1}{3}} \quad (15)$$

In the MIT bag model the speed of sound is a constant but in the VDW model the speed of sound depends upon the temperature, particle mass, and density of the system. There is a causality limit to the temperature dependent speed of sound at the speed of light indicated in Figure 2. To compare these values and contrast the VDW case with the ideal gas law case, we plot the ratio of the speed of sound to the MIT bag model in Figure 2 below.

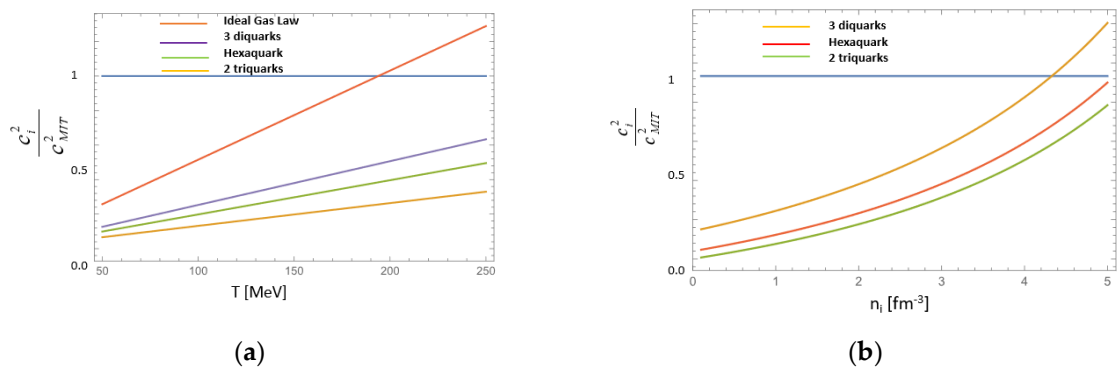

(a) (b)

**Figure 2.** (**a**) The variation in the speed of sound squared with temperature for the 3- diquark, hexaquark, and 2- triquark cases for the VDW and the ideal gas law EoS compared to the MIT bag model. (**b**) The variation in the speed of sound with density for the hexaquark, diquark and triquark cases for the VDW and ideal gas law EoS compared to the MIT bag model.

These values can now be used as indicators of potential quark states inside a compact stellar core of sufficient density where quark clustering, quasi-parton formation, mixed states, and phase changes can play a role in the possible final state configurations.

## 3. Stellar TOV Equations and the Chemical Potential

To investigate the conditions where a chemical potential phase transition can occur, we consider a static spherically symmetric mass as a dense stellar core with an ideal fluid source using the metric ansatz

$$ds^2 = g_{\mu\nu}dx^\mu dx^\nu = e^{-2\Phi(r)}dt^2 + \left(1 - \frac{2m(r)}{r}\right)dr^2 + r^2 d\theta^2 + r^2 \sin^2\theta d\phi^2 \quad (16)$$

with the Einstein and stress energy tensors given as

$$G_{\mu\nu} = R_{\mu\nu} - \frac{1}{2}g_{\mu\nu}R = 8\pi T_{\mu\nu}$$
$$T_{\mu\nu} = (\varepsilon + P)u_\mu u_\nu - g_{\mu\nu}P \quad (17)$$

The resulting TOV equations for the pressure and mass are

$$\frac{dp}{dr} = \frac{(\rho(r) + p(r))[m(r) + 4\pi r^3 p(r)]}{r[r - 2m(r)]}$$
$$\frac{dm(r)}{dr} = 4\pi r^2 \rho(r) \quad (18)$$

which, when combined with the EoS, provide a system of equations describing the stellar core. Here we follow Hajizadeh [70] and change variables from pressure and energy density, $\varepsilon$, to the chemical potential, $\mu$, and express the pressure equation in terms of the total baryon chemical potential as

$$\frac{d\mu}{dr} = \frac{\mu(r)[m(r) + 4\pi r^3 p(r)]}{r[r - 2m(r)]} \quad (19)$$

to examine the radial dependence in the interior of the stellar core. These equations represent a system of equations that can be solved numerically; however, to compare to the MIT Bag model, there is a stability requirement on the bag constant when strange matter is present. We will utilize the Franzon stability requirement to constrain values of the bag constant, $30$ MeV/fm³ $< B < 75.5$ MeV/fm³, to compare the MIT bag model to the VDW model. There is also a constraint on the maximum mass for a given radius, requiring the core to not form a Schwarzschild black hole (R < 2 M or, in terms of compactness, C=M/R < 0.3). To incorporate the bag constant constraint, we solve the TOV equations with the MIT bag model EoS at the two limits and then identify a chemical potential value that gives the same mass radius curve. These curves are identified in Figure 3.

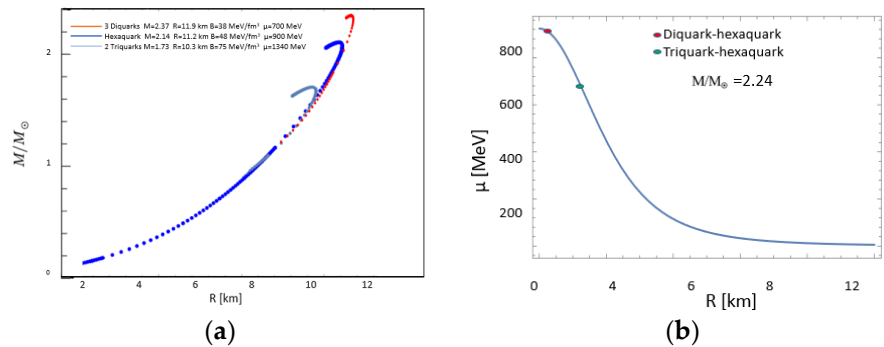

**Figure 3.** Chemical Potential vs radius inside a TOV star. In (**a**) the chemical potential as a function of radius is given for a 2 M star with the hexaquark threshold lines with the hexaquark, diquark, and mixed states shown. In (**b**) we numerically solve the chemical potential-modified TOV

equations, Equations (18) and (19), showing the maximum mass for the hexaquark, diquark, and triquark systems and indicating their respective chemical potentials.

Solving the TOV equations numerically for a sample stellar core results in a class of curves similar to the one shown in Figure 3. Likewise, plotting the mass-radius parametrization gives limiting curves, as seen in Figure 3, where three curves exhibiting the maximum mass and radius are given the MIT bag model constraints. The hexaquark chemical potential values for the mixed states indicate a more complex phase-change structure, especially for the non-color-singlet states.

## 4. Conclusions

We have investigated the chemical potential of the hexaquark using a phase changing multicomponent van der Waals equation of state within the context of high-density nuclear matter in the core of a cold beta equilibrium system. In particular, we have examined different internal quark clustering models of the hexaquark that involve diquark and triquark states that may not be in a color-singlet combination and simple mixtures of these states. This leads to different color factors to describe the different color force interaction strengths. Such differences can be approximately modeled in the multicomponent VDW EoS by using the mixing parameter for pairwise interactions that obey the VDW mixing rule. As a result, different hexaquark internal structure arrangements will give different values for the magnitude of the chemical potential of the hexaquark as shown in Table 4.

**Table 4.** The maximum mass and radius values that give the maximum compactness for different quark combinations corresponding to hexaquark internal structure with chemical potentials from Equation (9), where a Schwarzschild black hole has a compactness of 0.5.

| Quark Configuration | Mass [$M/M_\odot$] | Radius [km] | Compactness |
| --- | --- | --- | --- |
| 3-diquarks | 2.05 | 11.9 | 0.172 |
| 2-triquarks | 2.09 | 10.3 | 0.202 |
| Hexaquarks | 2.16 | 11.2 | 0.192 |
| Diquark:Hexaquark $k_{12}$=0.5 1:1 | 1.64 | 12.1 | 0.136 |
| $k_{12}$ = 0.5 2:1 | 1.58 | 12.8 | 0.078 |
| Triquark:Hexaquark $k_{12}$=0.5 1:1 | 1.73 | 11.6 | 0.149 |
| $k_{12}$=0.5 2:1 | 1.85 | 11.3 | 0.164 |

Our results are similar to Eduardo [71] and Kang [72], where they examined QCD EoS in compact stellar cores and analyzed chiral chemical potential limits associated with maxima in stellar mass values. Our values are closer to the chemical potential of Lopes in the Maxwell construction, assuming charge neutrality to establish lepton number densities for the stable values of the bag constant [73]. Here we find that, in comparison, the VDW model overestimates the magnitude of the chemical potential but does give a maximum limit for the hexaquark case, indicating they could form in a specific type of compact stellar core. In the TOV representation with a VDW EoS of a compact core, stability bounds, causality limits, and black hole formation all constrain the range of the chemical potential or induce phase changes for hexaquarks that would result in a layering of the core. Our key result for the VDW EoS in the TOV framework with a cold beta equilibrium system is that, for chemical potentials 700 MeV > $\mu$ > 1340 MeV with 1.73 > $M/M_{Solar}$ > 2.37 and 10.3 km > R > 11.9 km, there is no single state pure hexaquark core that remains stable without a phase transition. However, the mixed states of correlated diquarks and triquarks can cluster to form layers of increasing chemical potential towards the center of the star. This analysis did not include any boson- or color-superconducting formation properties which would soften the EoS in a fashion similar to a phase transition, but which allow for much higher chemical potentials ($\mu$ > 1400 MeV [74]) and include features we are now exploring. As higher resolution multispectral observations improve, it will soon be possible to begin to determine the nature of the interior of high-density neutron stars and or quark star candidates such as GW 170817 [75].

**Author Contributions:** Formal analysis: K.A., E.V.S., and K.A.A.; methodology: K.A., E.V.S., and K.A.A.; investigation: K.A., E.V.S., and K.A.A.; final analysis, numerical solutions, and visualization: K.A., E.V.S., and K.A.A.; writing—original Draft: K.A., E.V.S., and K.A.A.; writing—review and editing: K.A., E.V.S., and K.A.A. All authors have read and agreed to the published version of the manuscript.

**Funding:** This research received no external funding.

**Data Availability Statement:** Not applicable.

**Acknowledgments**: The authors wish to thank Western Kentucky University, Schlarman Academy for their kind support throughout the work on this project. All plots and numerical solutions were generated in Mathematica and the insightful and helpful comments from anonymous referees.

**Conflicts of Interest:** The authors declare no conflict of interest.